\documentclass[%
 reprint,
preprintnumbers,
nofootinbib,
 amsmath,amssymb,
 aps,
 prl,
]{revtex4-2}

\usepackage{graphicx}
\usepackage{dcolumn}
\usepackage{bm}
\usepackage{xspace}
\usepackage{mathrsfs}

\newcommand{\x}{\boldsymbol{x}}
\newcommand{\z}{\boldsymbol{z}}
\newcommand{\Exp}{\mathbb{E}\!}
\newcommand{\ExpSM}{\mathbb{E}_{\text{SM}}\!}
\newcommand{\blambda}{\boldsymbol{\lambda}}

\newcommand{\bkappa}{\boldsymbol{\kappa}}
\newcommand{\bkappahat}{\hat{\boldsymbol{\kappa}}}
\newcommand{\kappal}{\kappa_{\blambda}}
\newcommand{\kappalp}{\kappa_{\blambda'}}
\newcommand{\sigmall}{\sigma_{\blambda \blambda}}
\newcommand{\sigmallp}{\sigma_{\blambda \blambda'}}
\newcommand{\bomega}{\boldsymbol{\omega}}
\newcommand{\omegall}{\omega_{\blambda \blambda}}
\newcommand{\omegallp}{\omega_{\blambda \blambda'}}
\newcommand{\bomegabar}{\bar{\boldsymbol{\omega}}}
\newcommand{\omegallbar}{\bar{\omega}_{\blambda \blambda}}
\newcommand{\omegallpbar}{\bar{\omega}_{\blambda \blambda'}}
\newcommand{\bomegahat}{\hat{\boldsymbol{\omega}}}

\renewcommand{\L}{\mathcal{L}}
\newcommand{\Lumi}{\mathscr{L}}
\newcommand{\D}{\mathcal{D}}
\newcommand{\Dfull}{\mathcal{D}_{\mathrm{full}}}
\newcommand{\PW}{\ensuremath{\text{W}}\xspace}
\newcommand{\PZ}{\ensuremath{\text{Z}}\xspace}
\newcommand{\MW}{\ensuremath{M_\PW}\xspace}
\newcommand{\MZ}{\ensuremath{M_\PZ}\xspace}

\renewcommand*{\d}{\mathop{}\!\mathrm{d}}

\newcommand{\Sherpa}{S\protect\scalebox{0.8}{HERPA}\xspace}

\begin{document}
\preprint{COMETA-2026-40}

\title{An optimal observable for polarization measurements}
\author{Erik Bachmann}
\affiliation{Institute of Nuclear and Particle Physics, Technische Universit{\"a}t Dresden, 01062 Dresden, Germany}
\author{Mareen Hoppe}
\affiliation{Institute of Nuclear and Particle Physics, Technische Universit{\"a}t Dresden, 01062 Dresden, Germany}

\date{\today}

\begin{abstract}
Vector-boson polarization provides a sensitive probe of new physics in the electroweak sector. Rare processes and suppressed longitudinal contributions make optimal polarization separation essential to exploit the available data. We show that parton-level differential polarization fractions define optimal observables and a fundamental sensitivity limit for polarized cross-section measurements. Their conditional expectations retain all experimentally accessible polarization information, yielding the best achievable sensitivity within the assumed model, and can be learned by neural networks for unbinned or binned inference. Measuring their distributions combines polarization sensitivity with the model independence of fiducial cross sections, probing deviations beyond constant rescalings of polarized contributions. We demonstrate the approach in inclusive ZZ production. 


\end{abstract}

\maketitle


\section{Introduction}
Following the discovery of the Higgs boson in 2012 and subsequent measurements of its properties~\cite{ATLAS:2012yve,CMS:2012qbp}, a primary objective of the remaining Large Hadron Collider~(LHC) measurement campaigns is to reveal the nature of electroweak symmetry breaking~(EWSB) and possible new physics effects affecting it~\cite{ATLAS:2025eii}. The longitudinal polarization state of massive vector bosons is a direct consequence of EWSB---studying its interactions therefore provides a sensitive probe of new physics in the electroweak sector. A prominent example is vector boson scattering~(VBS), where delicate cancellations between Higgs and gauge contributions ensure perturbative unitarity in longitudinal scattering at high energies. New physics affecting EWSB might modify these cancellations and thereby alter the longitudinal contribution. Since this contribution is generally suppressed in already rare processes, the resulting deviations can be obscured in unpolarized measurements, making dedicated polarization measurements essential for their discovery.

Current state-of-the-art measurements perform binned template likelihood fits using polarized Monte Carlo~(MC) predictions to extract polarized cross sections or polarization fractions~\cite{ATLAS:2022oge, ATLAS:2024qbd,ATLAS:2023zrv,ATLAS:2026fhf,CMS:2020etf,ATLAS:2025wuw,ATLAS:2026kzu}, which require an observable that separates the polarized contributions. These measurements are statistically limited, with much larger contributions from other polarization states obscuring the longitudinal signal. For a given dataset, the choice of polarization-sensitive observable is therefore the central factor determining the measurement sensitivity. With the current strategy, projections for the measurement of the double-longitudinal state in VBS processes barely reach observation at the end of the HL-LHC data-taking campaign~\cite{ATLAS:2025eii, stange_2025_weap2-wms55}, highlighting the need for optimal polarization separation.

A classical approach to optimal discrimination between two hypotheses is the matrix element method~(MEM)~\cite{Kondo:1988yd,volobouev:2011,Fiedler:2010sg}. At parton level, it exploits ratios of squared matrix elements as likelihood-ratio discriminants, whose optimality follows from the Neyman--Pearson lemma. In this paper, we demonstrate that, in the spirit of the MEM, parton-level differential polarization fractions contain the full information needed to measure polarized cross sections or fractions and define a fundamental limit to the sensitivity of polarization measurements. Even after restriction to experimentally accessible information, their conditional expectations remain globally optimal within the model assumed by the cross section measurements and can be learned by a neural network regression. This provides a practical route to optimal observables that retain the full accessible polarization information and can be used directly in unbinned inference or binned for template fits. Furthermore, we propose a differential measurement of the learned polarization fractions as a novel measurement technique that combines the sensitivity of a polarization measurement with the model-independence of a fiducial cross section measurement. We demonstrate the approach in inclusive ZZ production, quantifying the sensitivity retained after restricting event information and binning, and illustrate the response of the learned observable to new-physics-induced shape modifications.

\section{Differential polarization fractions as optimal observable}
\subsection{The basis: differential, event-wise polarization fractions}
In this section, we discuss the definition of differential event-wise polarization fractions constituting the proposed observable. This is based on the method introduced in Ref.~\cite{Ballestrero:2017bxn} that has already been successfully applied in a plethora of phenomenological studies~\cite{Pellen:2021vpi,Denner:2020eck,Denner:2020bcz,Denner:2021csi,Denner:2022riz,Denner:2023ehn, Dao:2023pkl, Dao:2023kwc,Dao:2024ffg, Le:2022lrp, Le:2022ppa, Poncelet:2021jmj, Carrivale:2025mjy, Pelliccioli:2025com,Ballestrero:2017bxn, Ballestrero:2019qoy, Ballestrero:2020qgv, Denner:2024tlu, Denner:2025xdz, Denner:2026vmf,Le:2026oux,Hoppe:2023uux,Pelliccioli:2023zpd} and experimental measurements~\cite{ATLAS:2022oge, ATLAS:2024qbd,ATLAS:2023zrv,ATLAS:2026fhf,CMS:2020etf,ATLAS:2025wuw,ATLAS:2026kzu}. 
The method starts with performing a diagram selection, identifying only diagrams where the vector bosons of interest are resonant and single polarized states can be identified. Combined with an on-shell approximation, e.g. given by the narrow-width~(NWA)~\cite{Richardson:2001df, Uhlemann:2008pm, Artoisenet:2012st} or the pole approximation~\cite{Aeppli:1993rs,Denner:2000bj,Denner:2005fg,Denner:2019vbn}, gauge-independence of the method is retained. Polarized predictions can then be obtained from an amplitude with $n$ intermediate resonant vector bosons with momenta~$q_1,\dots, q_n$, masses~$m_1, \dots m_n$ and decay widths~$\Gamma_1, \dots \Gamma_n$:
\begin{align}
    \mathcal{M} = \mathcal{M}^{\mathcal{P}}_{\mu_1, \mu_2 \ldots} \left\{\prod^n_{j=1} \frac{i\left(-g^{\mu_j \nu_j} + \frac{q^{\mu_j}_j q^{\nu_j}_j}{m_j^2}\right) }{q_j^2-m_j^2+i\Gamma_jm_j} 
    \mathcal{M}^{\mathcal{D}}_{\nu_j}\right\},
\end{align}
where $\mathcal{M}^{\mathcal{P}}_{\mu_1, \mu_2 \ldots}$ denotes the production, and $\mathcal{M}^{\mathcal{D}}_{\nu_j}$ the decay part of the amplitude. Employing the completeness relation with polarization vectors~$\varepsilon^\mu_\lambda(q)$:
\begin{equation}
    \left(-g^{\mu \nu} + \frac{q^\mu q^\nu}{m^2}\right) = \sum_{\lambda=1}^3 \varepsilon^\mu_\lambda(q) \varepsilon^{\nu*}_\lambda(q),
    \label{Eq.: Completeness relation}
\end{equation}
and omitting the polarization sum yields the polarized amplitude:
\begin{equation}
        \mathcal{M}_{\lambda_1, \lambda_2, \ldots} = \mathcal{M}^{\mathcal{P}}_{\mu_1, \mu_2 \ldots} \left\{\prod^n_{j=1} \frac{i\,\varepsilon^{\mu_j}_{\lambda_j}(q)\varepsilon^{\nu_j*}_{\lambda_j}(q)}{q_j^2-m_j^2+i\Gamma_jm_j} \mathcal{M}^{\mathcal{D}}_{\nu_j}\right\}.
\end{equation}
Note that the size of the different contributions strongly depends on the chosen spin basis. The polarization index~$\lambda$ can take the values \{+,-,0\} which we call right, left and longitudinal polarization.

As Eq.~\eqref{Eq.: Completeness relation} already suggests, the overall squared amplitude for any given phase-space contribution normally receives contributions from several different polarization states. Their relative contribution to the unpolarized squared amplitude defines event-wise polarization fractions given by 
\begin{equation}
    \bomega \equiv \omega_{\blambda \blambda'} = \frac{\mathcal{M}_{\blambda} \mathcal{M}^*_{\blambda'}}{|\mathcal{M}_{\text{unpol}}|^2}, \,\,\,\, \text{where} \qquad \sum_{\blambda, \blambda'} \omega_{\blambda \blambda'} =1,
    \label{Eq.: pol fractions}
\end{equation}
with the multi-index~$\blambda = (\lambda_1, \lambda_2, ...)$. Note that PDF, phase space as well as flux factors are all identical between unpolarized and polarized cross sections, so defining $\bomega$ in terms of cross sections leads to the exact same result. 
Due to the polarization sum at amplitude-level in Eq.~\eqref{Eq.: Completeness relation}, we have to distinguish two types of polarization fractions if we inspect squared amplitudes:
\begin{align}
    |\mathcal{M}_{\text{unpol}}|^2 \,=\, \underbrace{\sum_{\blambda}  |\mathcal{M}_{\blambda}|^2}_{\text{pure polarization states}} + \underbrace{\sum_{\substack{\blambda, \blambda'\\\blambda \neq \blambda'}}  \mathcal{M}_{\blambda} \mathcal{M}^*_{\blambda'}}_{\text{interference}} . 
\end{align}
One set of fractions is formed from pure polarization states and one set contains the interferences between different polarization states.

\subsection{Polarization fractions as sufficient statistic}
As mentioned above, state-of-the-art polarization measurements rely on template-based maximum likelihood fits. In these fits, polarized MC predictions are used to construct polarization templates, i.e. distributions of the expected event yields as a function of a chosen observable~$\hat{O}$, that are fitted against unpolarized data to extract polarized cross sections or polarization fractions~\cite{ATLAS:2022oge, ATLAS:2024qbd,ATLAS:2023zrv,ATLAS:2026fhf,CMS:2020etf,ATLAS:2025wuw,ATLAS:2026kzu}. In many unpolarized measurements, the normalization of background contributions is only allowed to vary within prior uncertainties, are constrained using dedicated control regions, or both. In polarization measurements, however, the normalization of all polarization components can float freely in the fit and are the parameters of interest. One can intuitively understand that in this case, the sensitivity of the fit depends primarily on how well the corresponding differential distributions in~$\hat{O}$ can be distinguished from one another: the more dissimilar the distributions, the more information each observation~$\hat{O}(\x)$ carries about the parameters of interest.

Suppose a measurement aims to measure all polarized cross sections, parametrized by constant polarization modifiers~${\bkappa = \kappal}$ that scale the polarized amplitudes ${\mathcal{M}_{\blambda} \rightarrow \kappal \mathcal{M}_{\blambda}}$, of a multi-boson production process from a set of $N$ observed events~${\Dfull = \left\{ \x_1, \dots, \x_N \right\}}$. Here, $\x$ denotes the complete parton-level event configuration required to evaluate the matrix element which we for now deliberately assume to be fully accessible.

We can write the probability of observing the dataset~$\Dfull$ given the polarization modifiers, i.e. the extended likelihood, as
\begin{equation}\label{eq:likelihood_general}
    \L(\bkappa | \Dfull) = \frac{\nu(\bkappa)^N e^{-\nu(\bkappa)}}{N!} \prod_{n=1}^N p(\x_n | \bkappa),
\end{equation}
with $\nu(\bkappa)$ the expected event yield and $p(\x | \bkappa)$ the probability of observing event~$\x$ given the polarization modifiers~$\bkappa$. We can identify $p(\x | \bkappa)$ as the modified normalized differential cross section~$(\frac{1}{\sigma} \frac{\d \sigma}{\d \x})(\bkappa)$, which we can expand into a sum of scaled polarized differential cross sections:\footnote{We make the common simplifying assumption that each polarized amplitude is modified by a real multiplicative factor. This corresponds to neglecting possible BSM-induced relative complex phases between the polarized amplitudes.}
\begin{equation}\label{eq:diff_xs_expand}
        \frac{\d \sigma}{\d \x}(\bkappa) = \sum_{\blambda} \kappal^2 \cdot \frac{\d \sigmall^{\text{SM}}}{\d \x} + 2 \mathop{\mathrm{Re}}\sum_{\blambda < \blambda'} \kappal \cdot \kappalp \cdot \frac{\d \sigmallp^{\text{SM}}}{\d \x}.
\end{equation}
Inserting this into Eq.~\eqref{eq:likelihood_general}, expanding each factor with the unpolarized differential cross section~$\frac{\d \sigma_{\text{SM}}}{\d \x}$ and substituting $\nu(\bkappa) = \sigma(\bkappa) \cdot \mathscr{L}$ using the integrated luminosity~$\mathscr{L}$ finally yields
\begin{equation}\label{eq:unbinned_likelihood}
    \L(\bkappa | \Dfull) = \frac{\mathscr{L}^N}{N!} \left[ \prod_{n=1}^N \frac{\d \sigma_{\text{SM}}}{\d \x_n} \right] e^{-\sigma(\bkappa) \cdot \mathscr{L}} \cdot \prod_{n=1}^N r(\bkappa ; \x_n),
\end{equation}
with
\begin{equation}
    r(\bkappa ; \x) = \sum_{\blambda} \kappal^2 \cdot \omegall(\x) + 2 \mathop{\mathrm{Re}} \sum_{\blambda < \blambda'} \kappal \cdot \kappalp \cdot \omegallp(\x).
\end{equation}
We find that the parameter-dependent part of the likelihood depends on each event only through the polarization fractions. For $d$ separately measured polarization states, only the $d$ diagonal entries of~$\bomega$ and the $d(d-1)/2$ real parts of its off-diagonal entries, one of which is fixed by normalization, enter the cross section ratio~$r$. States not measured separately---e.g. $+$ and $-$ combined into a transverse state---share a common modifier, so their fractions, including pairwise interferences, enter only through their sum. Therefore, according to the Fisher--Neyman factorization theorem \cite{book:Casella_Berger_2024}, the vector of these independent components of~$\bomega$ is a sufficient event-level statistic for~$\bkappa$. It contains all information each event provides about the unknown polarization modifiers, making it an optimal observable for measuring polarized cross sections or fractions. Consequently, it can be used to derive a fundamental limit on the achievable polarization sensitivity; further details can be found in the appendix.

\section{Optimal observable with incomplete event information}
\subsection{Sufficiency after information loss}
Until now we have considered an idealized measurement with complete access to the full parton-level event information~$\x$. For an experimentally observed event~$\z$, however, parts of the event kinematics, such as initial-state momenta and parton flavors, are inaccessible. Moreover, effects beyond the parton level including parton showering and detector effects can introduce ambiguities and lead to a further loss of information.

We describe the mapping from the parton-level event configuration~$\x$ to the observed event~$\z$ by a measurement response~$T(\z|\x)$, which we assume to be independent of the polarization modifiers~$\bkappa$. The observed differential cross section is then given by marginalizing over the unobserved event information:
\begin{equation}
    \frac{\d \sigma}{\d \z}(\bkappa) = \int T(\z|\x) \frac{\d \sigma_{\text{SM}}}{\d \x} r(\bkappa ; \x) \d \x.
\end{equation}
Dividing by the corresponding SM differential cross section gives
\begin{align}\label{eq:r_reco}
    r(\bkappa ; \z) &\equiv \frac{\d \sigma(\bkappa)/\d\z}{\d \sigma_{\mathrm{SM}}/\d\z} =\frac{\int T(\z|\x)\frac{\d \sigma_{\mathrm{SM}}}{\d \x}r(\bkappa ; \x) \d\x}{\int T(\z|\x)\frac{\d \sigma_{\mathrm{SM}}}{\d \x} \d\x}\nonumber\\
    &=\sum_{\blambda}\kappal^2 \, \omegallbar(\z)+2\operatorname{Re}\sum_{\blambda<\blambda'}\kappal\kappalp \, \omegallpbar(\z),
\end{align}
where we have introduced the conditional expected polarization fractions:
\begin{equation}\label{eq:conditional_fractions}
    \omegallpbar(\z) \equiv \ExpSM \left[ \omegallp(\x) | \z \right].
\end{equation}
Crucially, marginalization over the inaccessible event information does not change the functional dependence on the polarization modifiers. It only replaces each parton-level polarization fraction by its conditional expectation given the observed event.

For a dataset of observed events~$\D = \{\z_1, \dots, \z_N\}$, the extended likelihood can consequently be written as
\begin{equation}
    \L(\bkappa|\D) = \frac{\Lumi^N}{N!} \left[ \prod_{n=1}^{N} \frac{\d \sigma_{\text{SM}}}{\d \z_n} \right] e^{-\sigma_{\text{obs}}(\bkappa) \Lumi} \prod_{n=1}^{N} r(\bkappa ; \z_n),
\end{equation}
where $\sigma_{\text{obs}}$ denotes the cross section in the observed phase space. We find that the dependence of the parameter-dependent part of the likelihood on each observed event enters exclusively through the conditional expected polarization fractions. By the Fisher--Neyman factorization theorem, the vector of independent components of~$\bomegabar(\z)$ is therefore a sufficient event-level statistic for the polarization modifiers~$\bkappa$. Consequently, this vector constitutes the optimal polarization observable that can be constructed from the experimentally accessible event information~$\z$---no other statistic constructed from~$\z$ can contain additional information on~$\bkappa$.

\subsection{Learning the conditional polarization fractions}
The conditional polarization fractions introduced above are defined by a marginalization over the unobserved event information. This is closely related to the MEM which evaluates event probabilities by integrating squared matrix elements against transfer functions~\cite{volobouev:2011}. For a realistic event simulation including effects beyond the parton level, this integration is generally not tractable event by event. Nevertheless, MC generators can provide paired samples $(\z, \bomega(\x))$ that allow learning the required conditional expectations by regression, a technique that has previously been used for polarization reweighting~\cite{Grossi:2023, Cruz-Martinez:2026}. Here, we establish that the regression converges to the sufficient statistic derived above.

Consider a neural-network prediction~$\bomegahat_\theta(\z)$ with parameters~$\theta$ trained on events distributed according to the SM using mean-squared-error loss:\footnote{Here, $\bomega$ denotes the vector of independent components defined above and $\|\cdot\|$ its Euclidean norm.}
\begin{align}
    R(\theta) &= \ExpSM \left[ \left\| \bomega(\x) - \bomegahat_\theta(\z) \right\|^2 \right] \nonumber \\
    &= \ExpSM \left[ \left\| \bomega(\x) - \bomegabar(\z) \right\|^2 \right] + \ExpSM \left[ \left\| \bomegabar(\z) - \bomegahat_\theta(\z) \right\|^2 \right]\!.
\end{align}
The first term is independent of $\theta$ and corresponds to the polarization information lost in the mapping~${\x \mapsto \z}$. Consequently, the population minimum is achieved at~$\bomegahat_\theta(\z) = \bomegabar(\z)$. Thus, in the limit of sufficient training statistics and model capacity, a regression of parton-level polarization fractions given the experimentally accessible event information converges precisely to the sufficient statistic derived above. 

This construction follows the simulation-based inference strategy of regressing accessible simulator information to obtain otherwise intractable likelihood quantities~\cite{Brehmer:2018}. Here, the polarization modifier parametrization offers an additional simplification: the exact quadratic dependence of the cross-section ratio in Eq.~\eqref{eq:r_reco} on~$\bkappa$ is preserved under marginalization, leaving only parameter-independent coefficient functions to be learned. A single regression trained on SM events therefore yields an experimentally accessible statistic that is globally sufficient throughout the modifier parameter space, without restricting the inference to small deviations from the SM.

For weighted MC simulation, the regression loss includes the event weights. Even for negative event weights as encountered in NLO QCD predictions, the population target is preserved, assuming the predicted unpolarized differential cross section is non-negative for all observed phase-space points. Further details and implications for the training are discussed in the appendix.

\section{Experimental inference}
Replacing the conditional fractions in Eq.~\eqref{eq:r_reco} by their predictions defines an estimated cross-section ratio~$\hat r(\bkappa;\z)$. The corresponding approximate extended likelihood, up to a parameter-independent factor, is
\begin{equation}
    \widehat{\L}(\bkappa | \D) \propto e^{-\Lumi \, \sigma_{\text{obs}}(\bkappa)} \prod_{n=1}^{N} \hat r(\bkappa;\z_n),
\end{equation}
and can be used unbinned for the statistical inference. In the limit of exact regression, this construction retains the full information each observation~$\z$ provides about the polarization modifiers. At finite accuracy, however, errors in the predicted coefficients can distort the likelihood and bias the inferred modifiers. Validation should therefore test whether the network has learned the required conditional expectations. Eq.~\eqref{eq:conditional_fractions} implies that for every observable~$O$:
\begin{equation}
    \ExpSM \! \left[\bigl(\bomegahat(\z)-\bomega(\x)\bigr)O(\z) \right]=0.
\end{equation}

A template fit provides a practical alternative in which the learned fractions only determine bin assignments, while polarization weights from MC simulation determine the expected bin yields~$\nu_b(\bkappa)$ as functions of the polarization modifiers. Since the network predictions do not replace these weights, imperfect regression can reduce sensitivity without itself invalidating the resulting Poisson likelihood.

The information lost in the binning procedure is governed by the within-bin variation of the event contribution to the extended-likelihood score~${\boldsymbol{s}(\z; \bkappa) = \nabla_{\bkappa} \ln r(\bkappa; \z)}$. As a special case of the missing-information principle~\cite{Orchard:1972,Louis:1982}, the Fisher information of the unbinned observed events and the binned counts differ by
\begin{equation}
    I_{\z} -I_{\text{bin}} =\sum_b \nu_b(\bkappa)\, \operatorname{Cov}_{\bkappa} \! \left[ \boldsymbol{s}(\z; \bkappa) \mid b \right],
\end{equation}
where conditioning on~$b$ restricts events to a given bin~$b$. This motivates binnings that group events with similar responses to small changes in the modifiers around a chosen reference parameter point. The specific binning can be selected using adaptive partitioning methods such as recursive splitting or iterative merging~\cite{Hastie:2009}, with an objective based on the binned Fisher information, such as the expected profiled uncertainty, or the Asimov test statistic for a discovery test.

This partitioning of the measurement phase space also sparks a completely orthogonal idea: instead of fitting polarized templates to data on detector-level, one can measure the unpolarized cross section in each sub-phase-space, performing essentially a differential measurement of our proposed observable. This approach retains the sensitivity of a template-based polarization measurement to modification of the polarized cross sections while gaining the model-independence of a fiducial differential cross section measurement. Furthermore, since our observable separates phase-space regions that are dominated by different polarizations by construction, its measurement can additionally probe non-trivial polarization modifications beyond flat cross section changes.

\section{Proof of concept: inclusive ZZ production}\label{Sec.: Conceptual proof}
As a proof of concept for our proposed observable, we consider inclusive ZZ production at the LHC, ${\mathrm{pp} \rightarrow \ell \bar{\ell} \ell' \bar{\ell'}}$ with ${\ell = \mathrm{e}^\pm, \mu^\pm}$, at 13.6 TeV. This process provides a non-trivial polarization structure with a clean signature and very low experimental backgrounds, making our signal-only model a reasonable approximation of the achievable measurement sensitivity. Predictions have been generated with the MC event generator \Sherpa~\cite{Sherpa:2019gpd, Sherpa:2024mfk} in version 3.0.4 in a MEPS@NLO setup~\cite{Gehrmann:2012yg,Hoeche:2009rj,Hoeche:2012yf,Hoeche:2014rya} with 0,1j@NLO QCD+2,3j@LO. It can provide cross-sections in a spin-correlated narrow-width approximation~\cite{Hoche:2014kca} for all polarization combinations as additional event weights~\cite{Hoppe:2023uux}, from which the polarization fractions as defined in Eq.~\eqref{Eq.: pol fractions} can be calculated. Polarization fractions are calculated in \Sherpa's nLO QCD approximation described in Ref.~\cite{Hoppe:2023uux} in the diboson center-of-mass frame. Final-state leptons are reconstructed following object reconstruction and phase-space selections of the most recent ATLAS measurement of the considered process~\cite{ATLAS:2026fhf}. Details on the simulation setup can be found in the Appendix.

We consider three pure polarization components---double-longitudinal (LL), double-transverse (TT) and the coherent sum of the two mixed longitudinal-transverse states (LT+TL)---as well as the corresponding three interference terms~(${d = 3}$). The corresponding real polarization modifiers are denoted by $\bkappa = (\kappa_{\mathrm{LL}}, \kappa_{\mathrm{TT}}, \kappa_{\mathrm{LT}})$, respectively. We impose $\kappa_{\mathrm{TT}} \geq 0$ to remove the global sign degeneracy, while allowing the other modifiers to take either sign.

\begin{figure}[tb]
    \centering
    \includegraphics[width=\linewidth]{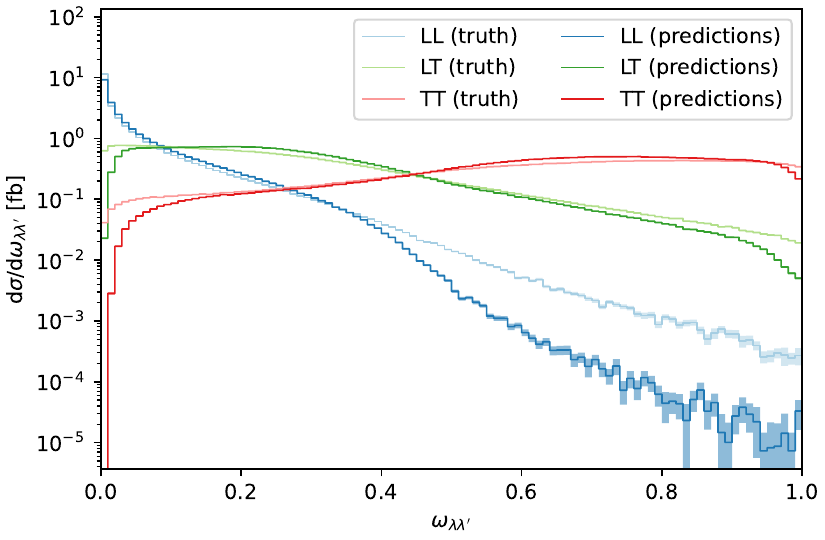}
    \caption{Differential cross sections as functions of the LL, LT+TL and TT polarization fractions in inclusive ZZ production. Parton-level fractions~(light curves) are compared with network predictions of their conditional expectations given the four-momenta and particle types of the four charged leptons~(dark curves).}
    \label{fig:learned_pol_fracs}
\end{figure}

We want to quantify how much polarization sensitivity is retained when restricting the full parton-level information to the four charged leptons and subsequently compressing the learned observable into template bins. The lepton quadruplet provides a clean, experimentally accessible event representation, although additional information may be available in the rest of the final state. While the parton-level polarization fractions~$\bomega(\x)$ are directly provided by \Sherpa, their conditional expectations~$\bomegabar(\z)$ are approximated by training a transformer-based neural network~\cite{vaswani2023attentionneed} to regress~$\bomega(\x)$ from the four-momenta and particle types of these leptons. Figure~\ref{fig:learned_pol_fracs} compares the parton-level and learned fractions; assuming ideal regression, their difference illustrates the effect of averaging over event information not included in the network inputs.

The expected sensitivity is evaluated at $\Lumi=164~\mathrm{fb}^{-1}$ using SM Asimov data, $\bkappa=\boldsymbol{1}$, for two complementary objectives: a simultaneous measurement of the three polarization modifiers and rejection of the hypothesis $\kappa_{\mathrm{LL}}=0$. Both analyses are performed using unbinned parton-level fractions as well as unbinned and binned network predictions. For the template fits, separate rectangular partitions are constructed by recursive splitting, maximizing the profiled Fisher information for $\kappa_{\mathrm{LL}}$ at the SM point for the measurement and the profiled null deviance for discovery. In all fits, backgrounds and systematic uncertainties are neglected.

\begin{table}[tb]
    \centering
    \caption{Expected sensitivity at $164~\mathrm{fb}^{-1}$ for SM Asimov data. Uncertainties are obtained from the HESSE covariance matrix of the simultaneous three-parameter fit. Significances use the asymptotic two-sided test of the signed LL modifier, expressed as one-sided Gaussian equivalents. The binned measurement and discovery results use separately optimized partitions.}
    \begin{ruledtabular}
        \begin{tabular}{lccc} \label{tab:fit_results}
             & Parton-level & NN & NN \\
             & unbinned & unbinned & binned \\[+0.03cm]
            \hline \\[-0.28cm]
            $\Delta\kappa_{\mathrm{LL}}$ & 0.0701 & 0.0804 & 0.0877 \\
            $\Delta\kappa_{\mathrm{LT}}$ & 0.0272 & 0.0311 & 0.0356 \\
            $\Delta\kappa_{\mathrm{TT}}$ & 0.0120 & 0.0129 & 0.0141 \\[+0.03cm]
            \hline \\[-0.28cm]
            $Z_{\mathrm{LL}}$ & 9.70 & 7.88 & 6.49 \\
        \end{tabular}
    \end{ruledtabular}
\end{table}

The results are summarized in Table~\ref{tab:fit_results}. At the SM point, the unbinned network predictions retain approximately 76\% of the profiled Fisher information for $\kappa_{\mathrm{LL}}$ available in the full parton-level polarization fractions, as inferred from the fitted uncertainties\footnote{For Asimov data, $\Delta\kappa_i = I_{i,\text{prof}}^{-1/2}$.}. A further 12\% of the full information is lost after binning the network predictions. The expected discovery significance decreases by similar amounts when replacing the parton-level fractions with network predictions and subsequently binning them, with relative reductions of approximately 19\% and 15\%, respectively. Overall, substantial sensitivity survives both the restriction to four-lepton information and the construction of a practical template observable.

\begin{figure}[tb]
    \centering
    \includegraphics[width=\linewidth]{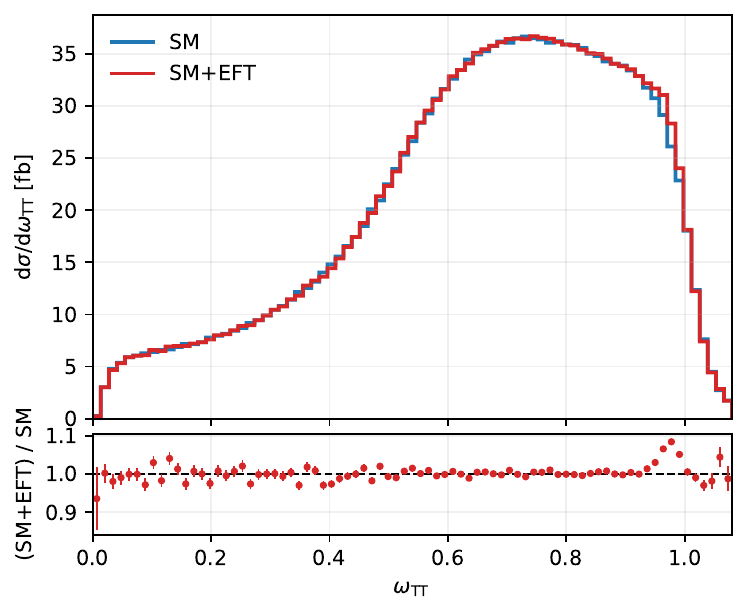}
    \caption{Differential cross sections as a function of the TT~polarization fraction for SM (blue) and SM+EFT predictions. The EFT contribution contains linear and quadratic terms for the $\mathcal{O}_{\tilde{B}W}$-operator with Wilson coefficient $C_{\tilde{B}W}/\Lambda^4$ of $2~\mathrm{TeV}^{-4}$.}
    \label{fig:smeft_omegaTT}
\end{figure}

To illustrate the potential impact of new physics on our observable, we consider neutral anomalous triple gauge couplings in dim-8 SMEFT~\cite{Degrande:2013kka} and generate unpolarized pseudodata events in NWA making use of \Sherpa's UFO interface~\cite{Hoche:2014kca, Degrande:2011ua,Darme:2023jdn} and the UFO model from Ref.~\cite{Degrande:2013kka}. We only study the CP-even $\mathcal{O}_{\tilde{B}W}=iH^\dagger \tilde{B}_{\mu\nu} W^{\mu \rho} \{D_\rho, D^\nu\} H+h.c.$ operator with a Wilson coefficient of $C_{\tilde{B}W}/\Lambda^4=2~\mathrm{TeV}^{-4}$ including linear and quadratic EFT terms. It introduces a ZZA vertex and only contributes to the production of one longitudinally and one transversally polarized boson~\cite{Degrande:2013kka}. Applying the neural network trained for Fig.~\ref{fig:learned_pol_fracs} on the generated pseudodata we obtain Fig.~\ref{fig:smeft_omegaTT} which shows the one-dimensional projection of our observable on the TT-axis defined in the SM. We can identify a clear excess in the EFT distribution of the observable around SM TT fraction of one illustrating the sensitivity of our observable to new physics effects. Apparently, triple gauge boson contributions described by the $\mathcal{O}_{\tilde{B}W}$ operators preferably populate phase-space regions where the bosons are almost entirely transversely polarized in the SM.  
\section{Conclusion}
In this letter, we have shown that event-wise polarization fractions are sufficient statistics for polarization modifiers that can be used to derive a fundamental sensitivity limit for polarization measurements. Their conditional expectations given experimentally accessible information remain sufficient and can be learned by neural-network regression, enabling unbinned inference or template fits. Measuring their distributions as fiducial differential cross sections additionally provides polarization-sensitive results without assuming constant polarization modifiers. An inclusive ZZ study was used as a proof of concept that quantifies the sensitivity retained after restricting event information and binning, and illustrates sensitivity to SMEFT-induced non-trivial polarization modifications.

\begin{acknowledgments}
\section{Acknowledgments}
EB and MH acknowledge the work of Lena Alshut published in Ref.~\cite{Alshut2026} conducting essential studies and tests towards this publication. EB and MH thank the Fermi National Accelerator Laboratory for hosting them during parts of the work. MH was financially supported by the German Academic Exchange Service~(DAAD) for that stay. Furthermore, MH acknowledges financial support from the Friedrich-Ebert-Stiftung during finalizing steps of this work. 

The authors acknowledge support from the COMETA COST Action CA22130 and thank Stefan Höche, Joany Manjarres, Matthieu Pellen, Giovanni Pelliccioli, Marek Schönherr and Frank Siegert for useful discussions regarding the work. Tim Herrmann, Michael Kobel, Frank Siegert are thanked for helpful comments to the manuscript.
\end{acknowledgments}

\bibliography{literature}

\appendix
\subsection{The fundamental limit of polarization sensitivity}
The sufficiency of~$\bomega$ implies that, given the full parton-level event kinematics~$\x$, no other observable can provide additional information about the polarization modifiers~$\bkappa$. Therefore, the corresponding full-information likelihood defines a fundamental limit on the polarization sensitivity achievable for the considered process, phase space and integrated luminosity. In the following, we quantify this limit for an idealized measurement with complete access to~$\x$.

For the previously introduced dataset~$\Dfull$, the likelihood ratio is given by
\begin{equation}
    Q(\bkappa) = -2 \ln \frac{\L(\bkappa | \Dfull)}{\L(\bkappahat | \Dfull)},
\end{equation}
with the maximum likelihood estimate~$\bkappahat$. To determine the median expected sensitivity, we need to evaluate its expectation for datasets generated under the SM hypothesis. For the extended unbinned likelihood, this leads to an unbinned Asimov construction in which the Poisson fluctuations of the total event yield and the distribution of event kinematics are replaced by their SM expectation value:
\begin{align}
    Q_A(\bkappa) = \ExpSM \left[ -2 \ln \frac{\L(\bkappa | \Dfull)}{\L(\boldsymbol{1} | \Dfull)} \right].
\end{align}
With $\frac{\d \sigma}{\d \x}(\bkappa) = \frac{\d \sigma_{\text{SM}}}{\d \x} r(\bkappa ; \x)$ and $\sigma(\bkappa) = \int \frac{\d \sigma}{\d \x}(\bkappa) \d \x$, inserting Eq.~\eqref{eq:unbinned_likelihood} yields:
\begin{align}
    Q_A(\bkappa) &= 2 \Lumi \int \frac{\d \sigma_{\text{SM}}}{\d \x} \left[ r(\bkappa ; \x) - 1 \right] \d \x \nonumber \\ 
      \qquad  & \qquad \qquad - 2 \ExpSM \left[ \sum_{n=1}^N \ln r(\bkappa ; \x_n) \right] \nonumber \\
    &= 2 \nu_{\text{SM}} \mathop{\ExpSM} \left[ r(\bkappa ; \x) - \ln r(\bkappa ; \x) - 1 \right]
\end{align}
with $\nu_{\text{SM}}$ the SM expected event yield. This quantity can be approximated using MC integration:
\begin{equation}
    Q_A(\bkappa) \simeq 2 \nu_{\text{SM}} \frac{\sum_n w_n\left[ r(\bkappa ; \x_n) - \ln r(\bkappa ; \x_n) - 1 \right]}{\sum_n w_n}.
\end{equation}
After profiling over all polarization modifiers other than the parameter of interest, the resulting profile-likelihood ratio can be used to determine the expected asymptotic confidence interval or discovery significance.

\section{Implications of signed event weights on the learned polarization fractions}
For weighted simulation, the weighted population loss is 
\begin{equation}
    R_w(\theta) = \Exp_{\,q} \! \left[w \left\| \bomega(\x) - \bomegahat_\theta(\z) \right\|^2 \right]
\end{equation}
with $q$ the sampling distribution of generated events and $w$ their event weights. It is minimized point-wise by
\begin{equation}\label{eq:regression_signed_optimum}
    \bomegahat_*(\z)= \frac{\Exp_{\, q} \! \left[ w \bomega \mid \z \right]}{\Exp_{\, q} \! \left[ w \mid \z \right]}
\end{equation} 
wherever $\Exp_{\, q}[w | \z] > 0$. Consequently, assuming the predicted SM differential cross section is positive at the considered observed phase-space point, the population minimizer equals the desired conditional polarization fractions; signed event weights by themselves do not change the population target. 

For a finite sample, a sufficiently flexible network may be able to isolate individual negative weight events and drive the empirical loss arbitrarily low by predicting arbitrarily large residuals. Regularization and early stopping can mitigate this instability, but the learned predictions must be assessed using independent validation samples. Replacing $w$ by $|w|$ would generally change the target.

\section{Simulation details}\label{Sec.: Simulation details}
This section summarizes some details on the MC predictions used for the presented proof-of-concept. As already mentioned in the main text, a multi-jet-merged event sample including fully-differential polarized cross sections is used. A merging scale of~$Q=20~\text{GeV}$ is assumed. Required tree-level matrix elements are obtained from \Sherpa's in-house matrix element generators COMIX~\cite{Gleisberg:2008fv} and AMEGIC~\cite{Krauss:2001iv}. One-loop amplitudes are provided by OPENLOOPS~\cite{Buccioni:2019sur,Buccioni:2017yxi,Cascioli:2011va}. Additional QCD and QED radiation are included, simulated by \Sherpa's default Catani-Seymour shower~\cite{Schumann:2007mg} and via YFS resummation~\cite{Yennie:1961ad,Schonherr:2008av,Flower:2022iew}, respectively. Non-perturbative QCD effects are incorporated by \Sherpa's default hadronization model~\cite{Winter:2003tt,Chahal:2022rid}, multiple-interaction simulation and hadron decay modeling. The PDF set \texttt{PDF4LHC21\_40\_pdfas}~\cite{PDF4LHCWorkingGroup:2022cjn} is accessed via the LHAPDF library~\cite{Buckley:2014ana}. The running of~$\alpha_S$ is extracted from the PDF set. We employ the $G_\mu$ scheme with $G_F=1.16638\cdot10^{-5}~\text{GeV}^{-2}$ and $\alpha_{G_\mu}\,=\,\frac{\sqrt{2}}{\pi}G_{\rm F}\MW^2\left(1-\frac{\MW^2}{\MZ^2}\right)$ assuming real masses and couplings for the massive vector bosons with pole masses and widths given by
\begin{equation}
\begin{aligned}
    m_\text{W}&=80.379~\text{GeV}, \qquad& \Gamma_\text{W}&=2.085~\text{GeV}, \\
    m_\text{Z}&=91.1876~\text{GeV}, \qquad& \Gamma_\text{Z}&=2.4952~\text{GeV}.    
\end{aligned}
\end{equation}
Factorization and renormalization scale are set to the Z boson pole mass.
Events are selected according to the following phase-space requirements
\begin{equation}
    \begin{aligned}
        p_{T,\,\mu^\pm}&>5~\mathrm{GeV}, \quad& |\eta_{\mu^\pm}|&<2.5,\\
        p_{T,\,\mathrm{e}^\pm}&>7~\text{GeV}, \quad& |\eta_{\mathrm{e}^\pm}|&<2.47,\\
        p_{T,\, l_1}&>27~\mathrm{GeV}, \quad& p_{T,\,l_2}&>20~\mathrm{GeV}, \\
        |m_{l^+l^-}-m_{\mathrm{Z}}|&<10~\mathrm{GeV}, \quad& \Delta R_{l^+l^-}&>0.05, \\
        m_{l^+l^-}&>5~\mathrm{GeV}, \quad& m_{4l}&>180~\mathrm{GeV}.
    \end{aligned}
\end{equation}
As input for the training of the neural network, we reconstruct leptons dressed with photons employing a cone dressing with resolution radius~$R=0.1$. Event selection and object reconstruction is performed by the help of RIVET 4~\cite{Bierlich:2024vqo}. The two opposite-sign-same-flavor lepton pairs with the smallest difference to on-shell bosons $\Delta m = |m_{l^+_1l^-_1} - m_{\mathrm{Z}}| + |m_{l^+_2l^-_2} - m_{\mathrm{Z}}|$ are identified as the Z boson decay products.
\end{document}